\documentclass[prb,twocolumn,showpacs,showkeys]{revtex4}%
\usepackage{graphicx}
\usepackage{amsmath}
\usepackage{amsfonts}
\usepackage{amssymb}

\newcommand{\Heb}{H_{eb}}
\newcommand{\Hc}{H_{c}}
\newcommand{\Hcool}{H_{cool}}

\newcommand{\Hsat}{H_{sat}}
\newcommand{\Jfm}{J_{FM}}
\newcommand{\Jaf}{J_{AF}}
\newcommand{\Jint}{J_{int}}
\newcommand{\Kfm}{K_{FM}}
\newcommand{\Kaf}{K_{AF}}

\newcommand{\Nc}{N_c}
\newcommand{\Maf}{M_{AF}}
\newcommand{\Ncif}{N_{cif}}
\newcommand{\Nshif}{N_{shif}}

\newcommand{\Nu}{N_{unc}}
\newcommand{\Rc}{R_c}
\newcommand{\tsh}{t_{sh}}
\date{\today}

\begin{document}
\title[Shape-dependent exchange bias effect]
{Shape-dependent exchange bias effect in magnetic nanoparticles with core-shell morphology}
\author{V.~Dimitriadis}
\author{D.~Kechrakos}
\email{dkehrakos@aspete.gr}
\affiliation{Department of Education, School of Pedagogical and Technological Education, Athens, GR-14121 }
\author{O.~Chubykalo-Fesenko}
\affiliation{Instituto de Ciencia de Materiales de Madrid, CSIC, Cantoblanco, Madrid, ES-28049}
\author{V.~Tsiantos}
\affiliation{Department of Electrical Engineering, East Macedonia and Thrace Institute of Technology, Kavala, GR-65404}
\begin{abstract}
We study the low-temperature isothermal magnetic hysteresis of cubical and spherical nanoparticles with ferromagnetic (FM) core - antiferromagnetic (AF)  shell morphology, in order to elucidate the sensitivity of the exchange bias effect to the shape of the particles and the structural imperfections at the core-shell interface. 
We model the magnetic structure using a classical Heisenberg Hamiltonian with uniaxial anisotropy and simulate the hysteresis loop using the Metropolis Monte Carlo algorithm. 
For nanoparticles with geometrically sharp interfaces, we find that cubes exhibit higher coercivity and lower exchange bias field than spheres of the same size. With increasing interface roughness, the shape-dependence of the characteristic fields gradually decays and eventually, the distinction between cubical and spherical particles is lost for moderately rough interfaces. 
The sensitivity of the exchange bias field to the microstructural details of the interface is quantified by a scaling factor ($b$) relating the bias field to the net moment of the AF shell $(\Heb=b\Maf+H_o)$. Cubical particles exhibit lower sensitivity to the dispersed values of the net interfacial moment.
\end{abstract}
\maketitle

\section{Introduction}

Magnetic nanoparticles composed of two coupled magnetic phases with distinct magnetic properties,  such as magnetic anisotropy, are presently extensively studied, owing to their potentials for a wide variety of  technological applications  ranging from magnetic recording \cite{sku03,eva13} to biomedicine.\cite{sta14} 
In particular, nanoparticles with the most commonly achieved core-shell morphology are grown either by vacuum chamber condensation methods \cite{zho07} or more often by colloidal chemistry methods.\cite{lop15}
The latter methods are low-cost and very versatile, producing nanoparticles with a very narrow size distribution and a variety of combined materials for the core and the shell. 
Owing to the flexibility offered by synthetic methods, nanoparticle shape has emerged as a promising tool for tailoring the magnetic properties at the nanoscale.\cite{khu13}
Prominent examples include engineering of nanoparticle shape to obtain a better response for the biomedical imaging  \cite{lee12} or  magnetic hyperthermia applications.\cite{bou13}
Nanoparticles with a ferromagnetic (FM)-core / antiferromagnetic (AF)-shell morphology are well known to exhibit the exchange bias effect after been field-cooled in a field.\cite{mei57,igl08}. 
Two characteristic features of the effect being the horizontal shift of the hysteresis loop in the direction opposite to the cooling field and the widening of the loop, expressed by an increase of the coercivity relative to the bare FM particle.  
In most cases, the exchange bias phenomenon arises from the pinning of magnetic moments at the interface between the two materials.\cite{nog05,igl08} Thus the material properties, the size and the atomic structure of the interface are crucial factors determining the existence and the strength of this phenomenon.  

Despite the abundance of experimental and theoretical works focusing on the role of the core size and shell size in controlling the exchange bias phenomenon in bi-phased nanoparticles \cite{igl08}, less attention has been payed so far to the interplay between exchange bias and particle shape. 
Experimental works have reported on the exchange bias effect in core-shell nanoparticles with cubical shape,\cite{zho07, sun11, pic11,noh12, khu15} 
and more recently nanoparticles with more complicated geometrical shape have been synthesized, including octahedra,  cuboctahedra and octopods,\cite{khu13} triangles,\cite{sha13, khu13} dimers and flowers\cite{cha13}, and their magnetic properties have been reported. 
At the same time, theoretical works on the magnetic properties of core-shell nanoparticles with non-spherical shape are a few,\cite{hu11,cha13} despite the necessity to unravel the interplay between exchange bias and particle shape. 
In addition to the role of particle shape, the importance of the FM-AF interface quality in controlling the magnetic properties of the exchange-coupled bi-phased nanoparticle has been recognized. 
In a recent work, Mao \textit{et al} \cite{mao12} demonstrated that the presence of non-magnetic  atoms (vacancies) at the core-shell interface degrades the exchange bias effect in FM/AF nanoparticles. 
Contrary to this trend, Evans \textit{et al} \cite{eva11} demonstrated that strong exchange bias can develop in weakly-coupled core-shell Co/CoO nanoparticles due to the presence of interface roughness. 
Finally, in recent experimental work by Juhin \textit{et al }\cite{juh14} direct evidence on the existence of an interdiffused layer at the interface of core-shell nanoparticles was provided, thus showing the way to obtaining atomic scale information on the quality of the buried core-shell interface. 

Provided that the overall shape of the nanoparticle and the atomic scale quality of the interface are two crucial factors determining the strength of the exchange bias effect, in the present work, we implement numerical simulations to study the interplay between these two factors in the case of cubical and spherical FM(core)-AFM(shell) nanoparticles. Our main conclusions are that cubes show higher coercivity and lower exchange bias field than spheres, while even moderate interface roughness washes out the shape-dependence of the exchange bias effect. Furthermore, in assemblies of cubical particles a wider dispersion of exchange bias field values is expected, due to the lower compensation of flat core-shell interfaces.   

\section{ Atomistic Model }

\begin{figure}[htb!]
\includegraphics[scale=0.25]{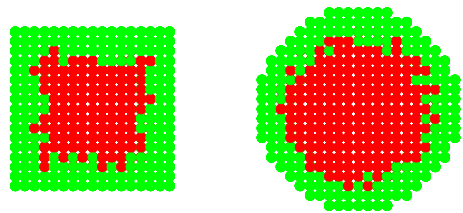}
\caption{(Color on line). Mid-plane of a cubical ($C58$) and a  spherical ($S710$) particle with rough interfaces. Roughness dispersion is $\sigma=0.5a$ in both cases.}
\label{f0}
\end{figure}

The atomic sites of a nanoparticle are generated by cutting a sphere of radius $R$ or a cube of edge $L$ from an infinite simple cubic (SC) lattice with lattice constant $a$. 
The center of the cube or the sphere coincides with the central atomic site of the nanoparticle. 
For both particle shapes, the core has the same geometrical shape as the whole particle and the shell thickness is taken $\tsh=3a$. 
The interface region  is considered two atomic layer thick, one atomic layer on either side. 
The structural parameters of the nanoparticles studied in the present work are summarized in Table \ref{tab1}. 
\begin{table}[Ht]   
\caption{Structural parameters of  modeled nanoparticles with ideal interfaces.
$R$ = radius (sphere) or half-edge (cube)
$R_c$ = core radius or core half-edge,
$N$=total number of spins (atoms),
$\Nc$=core spins,
$\Ncif$=core interface spins,
$\Nshif$=shell interface spins,
$\Nu$=uncompensated spins,
S=sphere, and C=cube.}
\begin{ruledtabular}
\begin{tabular}{ l l l l l l l l}            
Particle &$R/a$ &$R_c/a$ &$N$ &$\Nc$ &$\Ncif$ &$\Nshif$ &$\Nu$\\[1.0ex]
\hline
S710         &10.5  &7.5  &4945  &1189  &602   &762   &42   \\
S58          &8.0   &5.0  &2109  &257   &258   &410   &30   \\
C58          &8.0   &5.0  &4913  &729   &602   &866   &2    \\
C58r$^\star$ &8.0   &5.0  &4505  &549   &694   &642   &38   \\
\end{tabular}
\end{ruledtabular}
($^\star$) Cube faces along the  (110), ($\overline{1}$10) and (001) planes.
\label{tab1}
\end{table}
The particular choice of sizes is made in order to emphasize the importance of shape effects through comparison of cubes and spheres with the same size ($S58, C58, C58r$) or approximately the same volume, i.e. number of atoms ($S710, C58, C58r$).
The atomic scale structure of the core-shell interface is not easily controlled experimentally. In the most common case of surface oxidized metallic nanoparticles\cite{nog05}, the interface profile is formed by a diffusion-like process and is expected to have a highly complex structure containing various types of structural and substitutional defects.
Here, we approximate the complex interface profile by a model  of atomic scale roughness at the core-shell interface, which forms by the interpenetration of the magnetic phases. 
For simplicity we assume that roughness is confined within the two atomic layers of the interface region. 
Numerically, the rough interface profile of a spherical particle is realized by sampling the local radius at each site of the core-interface region from a Gaussian distribution with mean value equal to the core radius ($\Rc$) and variance $\sigma$. 
Correspondingly, for cubical particles, the local edge size is Gaussian distributed. Typical particles with rough interfaces are drawn in Fig.\ref{f0}
This atomistic model of interface disorder minimizes the formation of isolated substitutional defect sites\cite{eva11} and preserves to a reasonable degree the spatial separation of the two magnetic phases.

The magnetic structure is described by a classical Heisenberg Hamiltonian with first nearest neighbor exchange interactions and  uniaxial anisotropy along the $z$-axis, namely
$ H = -\sum_{<i,j>} J_{ij}\widehat{S}_i \cdot \widehat{S}_j
         -\sum_{i} K_i S_{iz}^2
         -H\sum_{i}S_{iz}$,
where $\widehat{S}_i$ is a classical spin vector on site $i$. 
The exchange energy constant  $J_{ij}$ takes the values $\Jfm$, $\Jaf$ and $\Jint$ depending on whether sites $i,j$ belong both to the FM region, both to the AF region, or one to the FM and the other to the AF region, respectively.
Similarly, the anisotropy energy constant $K_i$ takes the values $\Kfm$ and $\Kaf$, depending on the type of material on site $i$.  
Magnetostatic (dipolar) energy terms are neglected in the previous Hamiltonian, because the size of nanoparticles studied here, is well below the single-domain limit.\cite{comment-1}
In our simulations we use dimensionless energy parameters that have been previously introduced\cite{igl08}   to describe Co(core)/CoO(shell) nanoparticles. 
In particular, we define $\Jfm$ as the unit of energy ($\Jfm=1$) and then $\Jaf=-0.5\Jfm, \Jint=-0.5\Jfm, \Kfm=0.1\Jfm$, and  $\Kaf=0.5\Jfm$. 
In this set of parameters the AF anisotropy value seems to be too large ($K_{AF} \approx J_{AF}$)\cite{igl08}. 
However, we should note that since the AF shell is very thin ($\tsh=3a$), one expects the existence of strong surface anisotropy, which justifies the above choice of parameters. 
In our dimensionless units, temperature $T$ is measured in units of $\Jfm/k_B$ and the magnetic field strength $H$ in units of $\Jfm/g\mu_B$, where $\mu_B$ is the Bohr magneton and $g$ the  Land$\acute{e}$ factor. 

It is well known\cite{mei57, nog05} that in order to observe shifted hysteresis loops, the composite nanoparticles have to be field cooled from a high temperature ($T \gg T_C$) to a low temperature ($T \ll T_N$) in a relatively weak field ($\Hcool \ll \Hsat$) prior to sweeping the applied magnetic field $(   -\Hcool \leq H \leq +\Hcool)$. 
At the end of the FC process the FM moments are aligned along the field direction while the net AF moment is frozen parallel ($\Jint>0$)  or antiparallel ($\Jint<0$) to the cooling field producing an exchange-bias field on the FM.
Due to the uniaxial symmetry of our model Hamiltonian and the strong anisotropy of the AF material, the exact spin configuration of the AF shell at the end of the FC process can be approximately described by an Ising-type AF configuration,  which is either type-I $(+-+-)$  or type-II $(-+-+)$. 
Adopting this approximation for the FC state of the AF,  we avoid the time-consuming simulation of the actual FC process. 
Then the hysteresis loop simulation starts using as the initial spin configuration of the AF, the one (type-I or II) that minimizes the total energy of the particle under the applied $\Hcool$ field \cite{eva11}. 
Tests runs have shown that this approximation to the FC state, provides configurationally average loops almost identical to those obtained by the exact calculation of the FC process.
Hysteresis loops are simulated using the standard Metropolis Monte Carlo algorithm with single spin updates and $10^4$ initial Monte Carlo steps per spin (MCSS) for thermalization followed by $10^4$ MCSS for thermal averaging.
Sampling over thermal disorder is performed every $\tau=10$ MCSS to minimize correlations between sampling points.
The percentage of accepted spin moves is kept close to $50\%$ by adjusting the width of the spin moves.
A field sweep rate $\Delta H/MCSS=10^{-5}$ is used, and is kept constant throughout the present work, in order to point out changes in the loop characteristics arising solely by the structural and morphological properties of the particles and not the sampling time.
For a given particle shape, size and dispersion of roughness, a configurational average over the quenched interface disorder is performed using an assembly of $N_p=50$ non-interacting particles. 

\section{ Numerical Results}

\begin{figure}[htb!]
\includegraphics[scale=0.65]{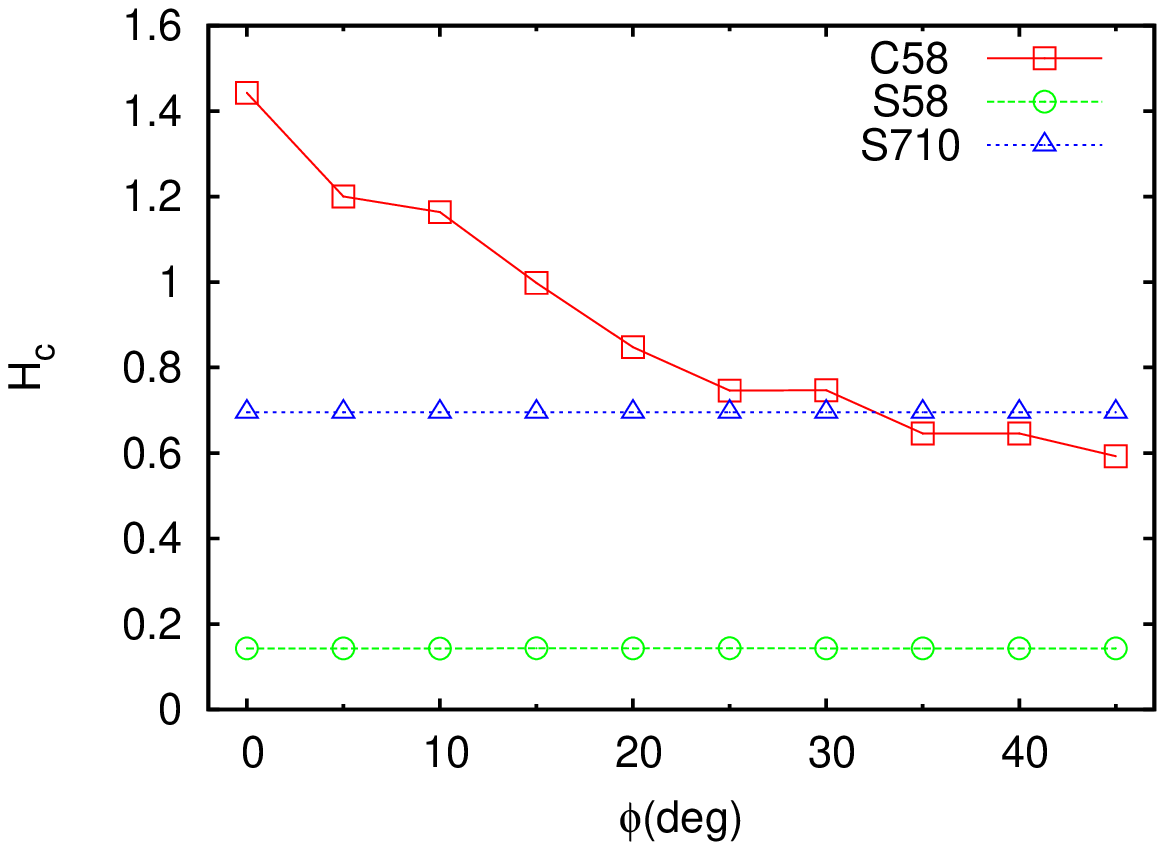}
\includegraphics[scale=0.65]{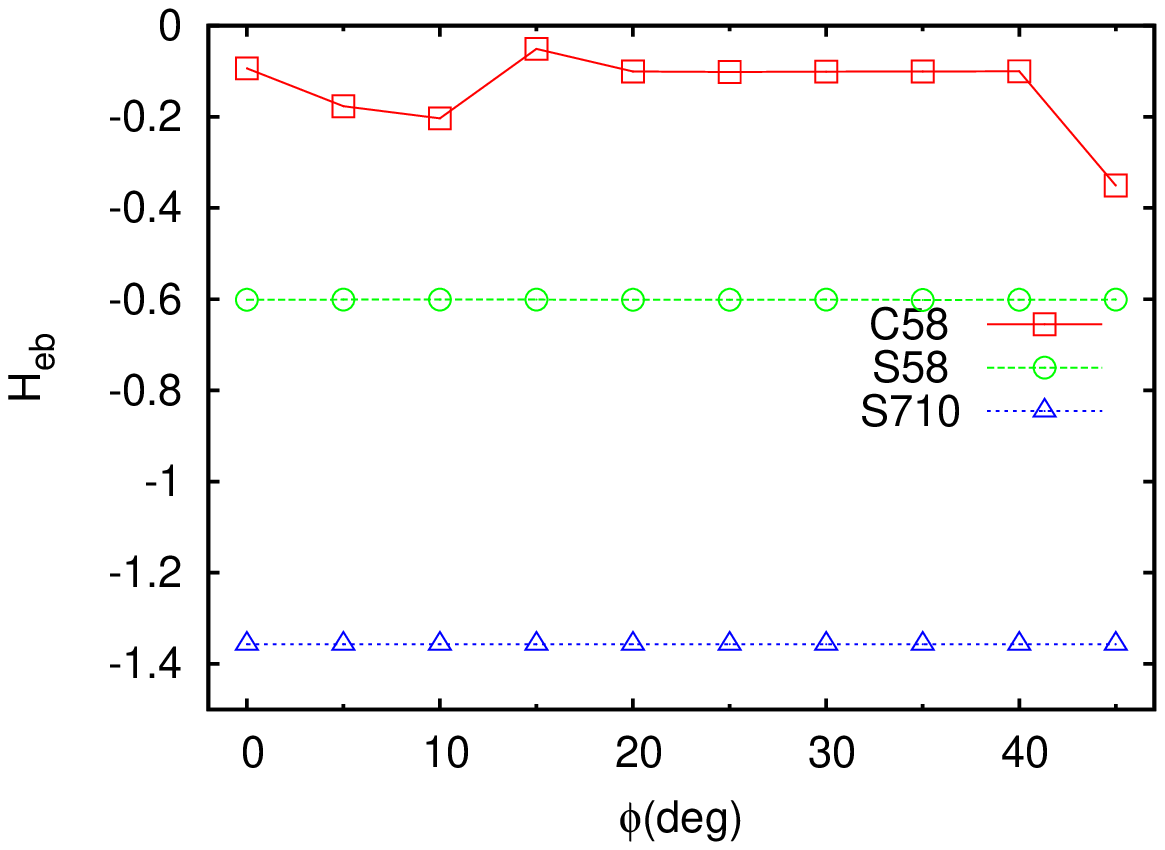}
\caption{(Color on line). Dependence of  low temperature ($T=10^{-3}$) coercivity (upper part) and exchange bias field (lower part) of cubical and spherical particles on the rotation angle ($\phi$) of the underlying SC lattice about the crystallographic axis [001].}
\label{f1}
\end{figure}

The origin of the exchange bias effect in FM/AFM nanostructures is attributed to the existence of uncompensated AF spins at the FM-AF interface\cite{nog05}.
For a spherical nanoparticle the interface is highly uncompensated due to its curved shape.
However, if a cubical particle is formed with faces parallel to the family \{100\} of crystallographic planes of a SC lattice, the core-shell interface is compensated (with exceptions at the corner sites), leading to a vanishing small exchange bias effect.
Uncompensated spins in a cubical core-shell particle exist provided the cube sides coincide with lower symmetry planes of the SC lattice. 
To achieve this we rotate the underlying SC lattice  by an arbitrary angle $\phi$ about the [001] axis. 
The dependence of $\Hc$ and $\Heb$ on the rotation angle $\phi$ is shown in Fig.\ref{f1}. Notice that for all $\phi$ angles, shown in  Fig.\ref{f1}, the applied field remains parallel to the easy axis of the nanoparticle. 
The coercivity and exchange bias field of the cubical particle depend substantially on the crystallographic orientation of the underlying lattice, while the corresponding fields of the spherical particle exhibit the expected rotational invariance (Fig.\ref{f1}).

Cubes with faces along \{100\} planes ($\phi=0$) have nearly twice the coercivity of  spheres of the same size. 
The mechanism leading to the enhanced coercivity of the FM core- AF shell nanoparticles is the 'drag' of shell-interface spins by the core-interface spins during magnetization reversal.\cite{nog05,igl08}. 
Consequently, the variation of $\Hc$ follows the variation in the number of dragged AF spins at the shell-interface ($\Nshif$), shown in Table \ref{tab1}.
The reduction of coercivity with rotation angle, as seen for $C58$, is a finite size effect. 
Namely, the dragging mechanism becomes less efficient as the fraction of dragged AF spins relative to the FM spins decreases. 
This situation occurs when the $\phi$-angle increases ($\Nshif/[\Nc+\Ncif]=0.651$ for $\phi=0^0$ and $0.516$ for $\phi=45^0$).

On the other hand, the exchange bias effect of the cubical particle increases (in absolute value) significantly for $\phi \approx 45^0$, approaching the value of $\Heb$ for a sphere with the same geometrical size.
This behavior is understood by the increase of the number of uncompensated spins of the rotated cube, which becomes comparable to the corresponding number for a sphere, as shown in Table \ref{tab1}.
Overall, the data in Fig.\ref{f1} demonstrate that for cubical and spherical particles with almost the same size (see Table \ref{tab1}), the same shell thickness and ideal FM-AF interfaces, cubical particles exhibit higher $\Hc$ values and lower $\Heb$  values than their spherical counterparts.
\begin{figure}[htb!]
\includegraphics[scale=0.65 ]{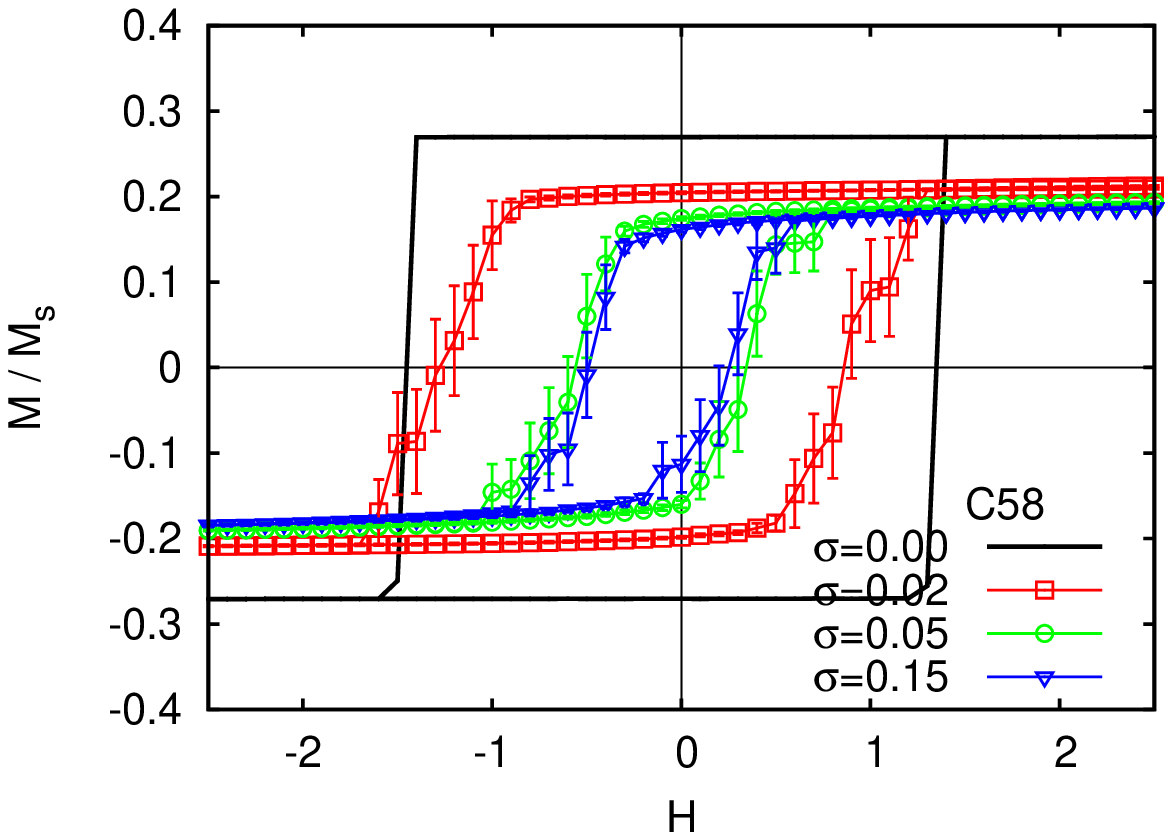}
\includegraphics[scale=0.65 ]{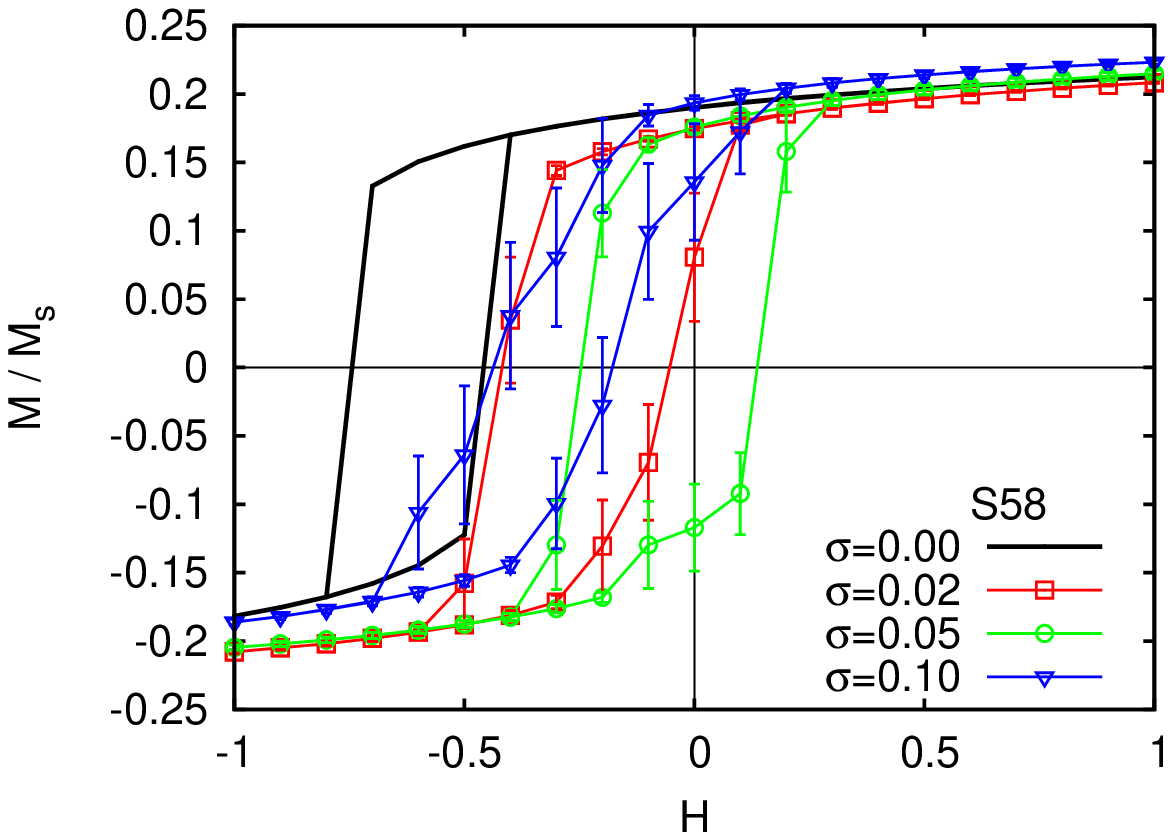}
\includegraphics[scale=0.65 ]{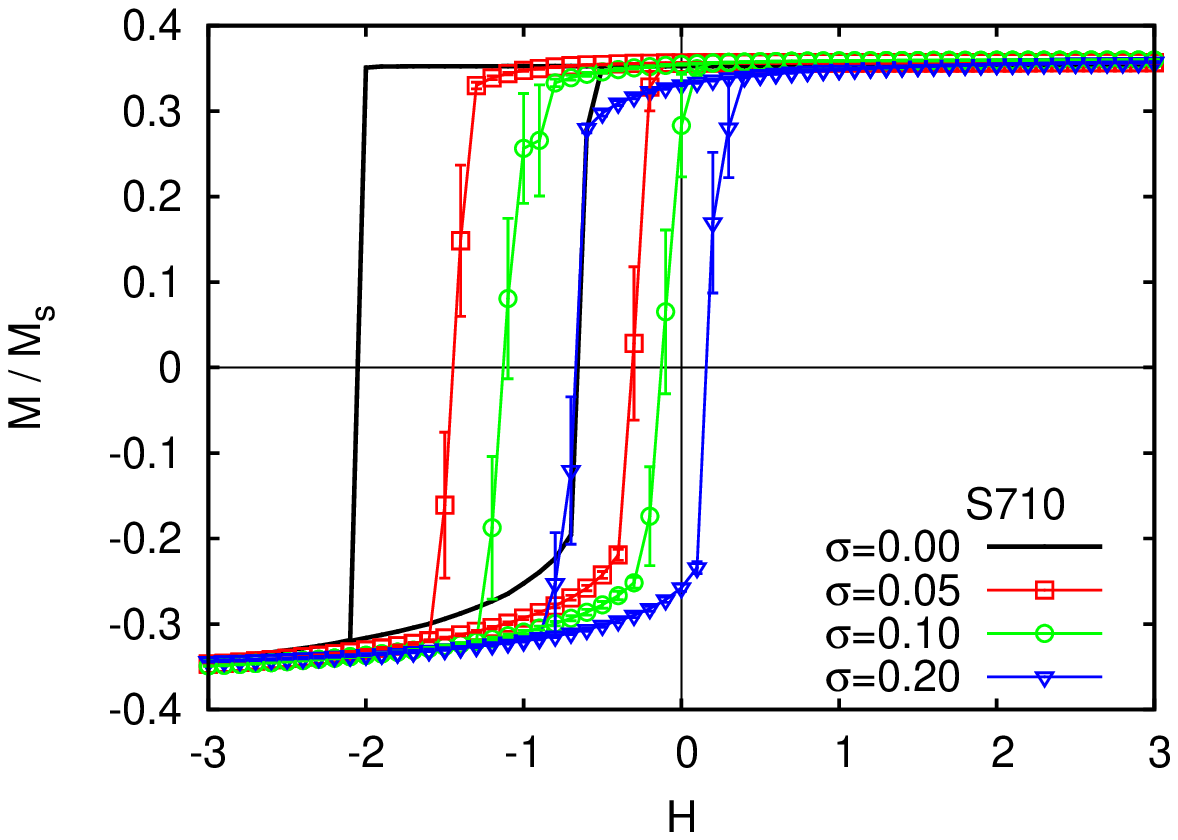}
\caption{(Color on line). Low-temperature ($T=10^{-3}$) hysteresis  loops of cubical  and  spherical particles with rough interfaces. Error bars arise from configurational averaging over the interface disorder.}
\label{f2}
\end{figure}

We consider next nanoparticles with roughness at the core-shell interface. 
In Fig.\ref{f2} we show the low-temperature ($T=10^{-3}$) hysteresis loops of cubes and spheres with rough interfaces.
The underlying SC lattice is not rotated ($\phi=0$) in these calculations, thus the loops depict the effect of roughness on the compensated interface of a cube ($C58$) and the uncompensated interface of the spheres ($S58, S710$).
In all cases, roughness causes loop shearing, due to the development of a wide barrier distribution at the interface region arising from the mixing of the two phases. 
Notice that the shearing effect occurs for the average loop, which represents the mean behavior of an assembly of particles with different realizations of interface disorder.
Individual loops (not shown here) have an almost rectangular shape, but they show a dispersion of $\Hc$ and $\Heb$ values, which leads upon averaging to the observed shearing effect of the average loop.\cite{eva11} 
Comparison of loops for $S58$ and $S710$ indicates that the shearing effect is stronger for the sphere with the smaller core, because of the increasing interface-to-volume ratio with decreasing core size.
\begin{figure}[htb!]
\includegraphics[scale=0.70]{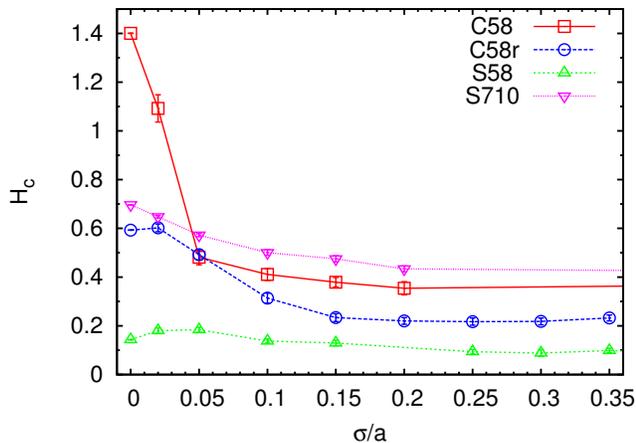}
\caption{(Color on line). Dependence of  coercivity  on interface roughness at low temperature ($T=10^{-3}$).}
\label{f3a}
\end{figure}

In Fig.\ref{f3a} we depict the dependence of $\Hc$ on interface roughness. For both cubes and spheres,  $\Hc$ shows a monotonous decrease with increasing roughness owing to the mixing of FM-AF at the interface that makes less effective the dragging of the shell interface moments by the core-interface moments during the magnetization reversal of the FM. A similar reduction of $\Hc$ with controlled interface roughness has been recently observed experimentally.\cite{juh14}.
Cubes display higher coercivity than spheres of the same size, $\Hc^{C58}>\Hc^{S58}$, even in the presence of roughness. 
Interesting, however, is the dramatic effect that even weak roughness can have on the coercivity. As seen in Fig.\ref{f3a}, the presence of weak roughness ($\sigma \approx 0.05a$) causes a dramatic drop of the coercivity of cube $C58$ below the coercivity of sphere $S710$. 
The sensitivity of $\Hc$ of the compensated cubical interface ($C58$) to disorder, implies that the experimentally measured coercivity values for cubes and spheres of similar sizes are screened by the presence of disorder. Detailed information on their interface structure\cite{juh14} would be extremely valuable before any conclusion correlating the coercivity values to the nanoparticle shape could be drawn.\cite{khu13, khu15} 
\begin{figure}[htb!]
\includegraphics[scale=0.65]{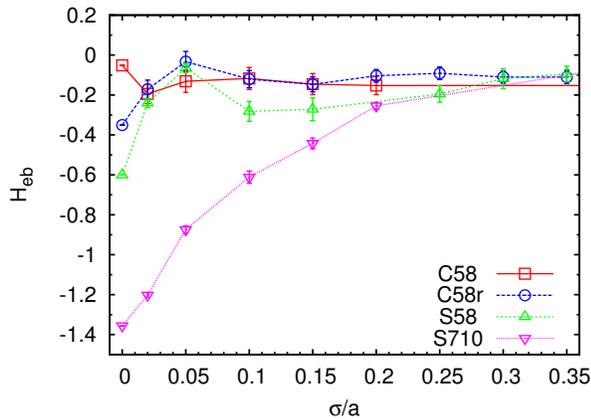}
\caption{(Color on line). Dependence of exchange bias field on interface roughness at low temperature ($T=10^{-3}$).}
\label{f3b}
\end{figure}

Consider next the effect of roughness on the exchange bias field, as depicted in Fig.\ref{f3b}.
Interface roughness can either enhance or suppress the values of the exchange bias field. 
Particle $C58$ shows increasing values of $|\Heb|$ with roughness, while $C58r$ and the spherical particles show decreasing values of $|\Heb|$.
The distinct effect of roughness on cubes and spheres is related to the compensation of their interfaces. 
In particular, the increase of $\Heb$ with roughness in $C58$ is in accordance to the predictions of the  random field model \cite{mal87}, namely that roughness on a flat compensated AF interface, creates uncompensated spins required for the loop to shift. Contrary to this trend, atomic scale roughness on uncompensated interfaces reduces the number of uncompensated spins leading to suppression of $\Heb$, as observed on the flat interface of $C58r$ and the curved interfaces of both spheres.
As the degree of roughness increases, the shape-dependent  $\Heb$, is gradually suppressed and cubes and spheres of  similar size have very similar values of $\Heb$ for $\sigma \geq 0.3$.
Furthermore, an interesting feature seen in Fig.\ref{f3b} is the pronounced minimum of $\Heb$ at $\sigma \approx 0.05$ for $C58r$ and $S58$. This stems form the wide dispersion of $\Heb$ values due to disorder that often assume positive values and produce on average a minimum in $|\Heb|$.  The non-monotonous dependence of $\Heb$ on roughness has been previously observed\cite{lei99} in exchanged-coupled bilayers and interpreted as the competition between interface exchange and Zeeman energy of the AF spins at the shell-interface, when the interface exchange is antiferromagnetic. We show here that a similar mechanism acts in core-shell nanoparticles irrespectively of their shape. 

\begin{figure}[htb!]
\includegraphics[scale=0.65 ]{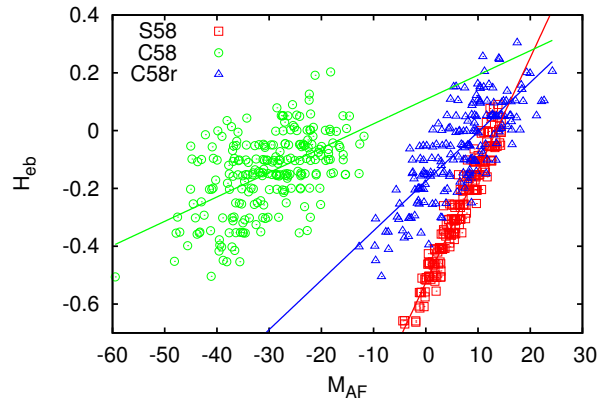}
\caption{ (Color on line). Scaling behavior of exchange bias field with the net magnetic moment of the AF shell for spherical and cubical nanoparticles with rough interfaces. Data in each case are selected from an assembly of $N_s$=150-200 non-interacting nanoparticles with interface roughness in the range $\sigma$ =0.05-0.20. 
The slopes of the least squares fits, shown by straight lines, are 
$b(S58)=0.039(\pm2\%)$,  
$b(C58)=0.008(\pm11\%)$ and 
$b(C58r)=0.017(\pm6\%)$ }  
\label{f4}
\end{figure}

The origin of the exchange bias effect in our model is the net (uncompensated) AF moment ($M_{AF}$) that is exchanged coupled to the FM.\cite{mei57}
Due to large anisotropy of the AF, which nearly freezes the net moment, the resulting exchange bias field acting on the FM is expected to obey a linear scaling law, namely $\Heb=b \dot M_{AF} + H_0$.  
When an assembly of nanoparticle is considered, $\Maf$ is expected to assume different values due to different realizations of the interface disorder. 
In this case, the scaling coefficient can be written as $b=d\Heb/dM_{AF}$, where the denominator expresses the variation of $\Maf$ values due to interface randomness. 
Therefore, one could interpret $b$ as the average (over a nanoparticle assembly) sensitivity of  $\Heb$ to the microstructural details of the interface.
In Fig. \ref{f4} we depict the values of $\Heb$ for an assembly of non-interacting particles as a function of the net moment of the AF shell at the FC state. 
The linear fit to the data provides the average sensitivity parameter ($b$), which is found to increase as the the uncompensation of the AF interface increases. 
It is notable also, that larger dispersion of $\Heb$ values occurs for cubes than for spheres, because interface structural imperfections constitute a stronger perturbation to compensated (cubic) interface, namely an assembly of cubes is characterized by larger dispersion of $\Heb$values than an assembly of spheres with the same size and degree of interface roughness. 

\section{ Conclusions}

In conclusion, our simulations of the exchange bias effect in  FM core - AF shell nanoparticles with cubical and spherical shape show that 
among particles with ideal (non-disordered) core-shell interfaces cubes have higher coercive field  ($\Hc^{cube}>\Hc^{sphere}$), owing to the larger number of 'dragged'  AF interface spins. This trend remains for uncompensated interfaces even in the presence of interface roughness. For compensated interfaces however, weak disorder can reverse this situation rendering spheres harder than cubes.
On the other hand, spheres exhibit higher exchange bias field ($\Heb^{sphere}>\Heb^{cube}$) due to the larger number of uncompensated spins on curved FM-AF interfaces. 
A non-monotonous dependence of $\Heb$ on roughness is observed, for small nanoparticles ($\Rc \leq 5a$) with antiferromagnetic interface exchange coupling and uncompensated interfaces, similar to the case of exchange coupled layered systems. Experiments on core-shell bi-magnetic nanoparticles with controlled interface roughness\cite{juh14} are required to investigate further this point. 
Strong roughness ($\sigma \ge 0.3$) suppresses the shape-dependence of $\Heb$ and $\Hc$. The atomistic details of the interface are washed out. The $\Heb$ nearly vanishes for both cubes and spheres and $\Hc$ retains a weak shape-dependence.
Shape-effects persist in the scaling factor relating the average exchange bias field of a (non-interacting) nanoparticle assembly to the net AF moment, characterized by  higher values for the spherical particles ($b^{sphere}>b^{cube}$). Assemblies of cubical particles are expected to have by a larger dispersion of $\Heb$ values than assemblies of spherical particles.

\begin{acknowledgments}
Research co-financed by the European Social Fund and Greek national funds through the Research Funding Program \textit{ARCHIMEDES-III}.
\end{acknowledgments}


\begin{thebibliography}{24}
\bibitem[{com()}]{comment-1}
\bibinfo{note}{The  magnetostatic (dipolar) energy of a nanoparticle is $E_{dip}  \approx \lambda g N^2$, with the dipolar strength $g=\mu_B^2/a^3$  and  $\lambda <1$ a numerical factor arising from the $1/r$ decay of dipolar interactions with distance between atomic sites. The exchange   energy is $E_{exc}\approx zJN$, with $z$ the coordination number. The two energy contributions are comparable when $N \approx zJ/ \lambda g \approx 10^6 $ or the critical particle diameter $L_c \approx N^{1/3}a \approx 100a$. For nanoparticles studied here ($L<<L_c$), dipolar interactions can be safely ignored.}
\end{thebibliography}
\end{document}